# Stokes-vector evolution in a weakly anisotropic inhomogeneous medium

Yu.A. Kravtsov[1,2], B. Bieg[1], and K.Yu. Bliokh[3,4,5,*]

[1]*Institute of Physics, Maritime University of Szczecin, 1-2 Waly Chrobrego St., Szczecin 70500, Poland*
[2]*Space Research Institute, Profsoyuznaya St. 82/34, Moscow 117997, Russia*
[3]*Institute of Radio Astronomy, 4 Krasnoznamyonnaya St., Kharkov 61002, Ukraine*
[4]*Optical Engineering Laboratory, Faculty of Mechanical Engineering, Technion–Israel Institute of Technology, Haifa 32000, Israel*
[5]*Amcrys Ltd., 60 Lenin Ave., Kharkov, 61001, Ukraine*
*Corresponding author: k_bliokh@mail.ru

Equation for evolution of the four-component Stokes vector in weakly anisotropic and smoothly inhomogeneous media is derived on the basis of quasi-isotropic approximation of the geometrical optics method, which provides consequent asymptotic solution of Maxwell equations. Our equation generalizes previous results, obtained for the normal propagation of electromagnetic waves in stratified media. It is valid for curvilinear rays with torsion and is capable to describe normal modes conversion in the inhomogeneous media. Remarkably, evolution of the four-component Stokes vector is described by the Bargmann–Michel–Telegdi equation for the relativistic spin precession, whereas the equation of the three-component Stokes vector resembles the Landau–Lifshitz equation for the spin precession in ferromagnetic systems. General theory is applied for analysis of polarization evolution in a magnetized plasma. We also emphasize fundamental features of the non-Abelian polarization evolution in anisotropic inhomogeneous media and illustrate them by simple examples.

## 1. Introduction

There are two main approaches for the description of electromagnetic wave propagation and polarization evolution in weakly anisotropic media. The first one deals with coupled wave equations for the components of the electromagnetic wave field. In stratified media it is the Budden's method [1–4], whereas the modification of Budden's approach for an arbitrarily inhomogeneous medium was suggested in [5] in the form of quasi-isotropic approximation (QIA) of the geometrical optics method. QIA was developed in depth in subsequent publications [6–9]. The second, alternative approach – "Stokes vector formalism" (SVF), initiated in papers [10,11] (see also [12]), was applied for the purposes of fiber [13] and plasma polarimetry [14–21]. Basically, it was developed for the simplest refractionless case of the normal propagation in a stratified medium; the effect of refraction is discussed briefly in [17]. This approach deals with evolution equation for the Stokes vector, and, thus, describes the polarization evolution explicitly in terms of standard parameters. In contrast to coupled wave equation approach, the Stokes vector formalism operates with quantities, which are quadratic in the field variables.

Comparative analysis of the two approaches mentioned, was performed by Segre in [19], where he has analyzed advantages and shortcomings of each technique, omitting their deep underlying equivalence. In fact, the QIA and SVF equations represent, respectively, differential Jones and Mueller calculus of the polarization evolution along the ray [11,13,20]. It was shown recently [22] that the evolution equation for three-component Stokes vector can be derived for the general case of arbitrarily inhomogeneous smooth refractive medium directly from Maxwell equations on the basis of formal quantum-mechanical approach. In such a framework, the QIA and SVF equations are just Schrodinger- and Heisenberg-type representations of the same polarization evolution equation, where the Stokes vector plays the role of the pseudo-spin in the problem.

The present paper intends to derive the evolution equation for the full (four-component) Stokes vector directly from quasi-isotropic approximation and thereby to reveal the equivalence of the two techniques under discussion. Our Stokes-vector equation generalizes previous results (obtained for the refractionless case) to the general case of arbitrarily inhomogeneous media and curvilinear rays. The paper is organized as follows. Section 2 presents the basic equations of quasi-isotropic approximation in different coordinate systems, including Popov's rotationless frame. Section 3 derives equations for four-component Stokes vector from QIA for weakly anisotropic medium of general type. Section 4 applies general theory for analysis of the Stokes vector evolution in a weakly anisotropic collisional plasma. Finally, Section 5 analyses fundamental features of the polarization evolution in weakly anisotropic inhomogeneous media (such as the normal mode conversion and non-Abelian nature of the polarization evolution) which are illustrated by simple examples.

## 2. Quasi-isotropic approximation

The geometrical optics method describes propagation of waves in smoothly inhomogeneous media. Let $k_0 = 2\pi/\lambda_0$ be the wave number in vacuum ($\lambda_0$ is the wavelength), whereas $L$ be the characteristic scale of the medium inhomogeneity. Applicability of the geometrical optics requires inequality

$$\mu_G = \frac{1}{k_0 L} = \frac{\lambda_0}{2\pi L} \ll 1, \tag{1}$$

where $\mu_G$ is the geometrical optics small parameter [8,9,23].

In a weakly anisotropic medium, dielectric tensor consists of the predominant isotropic permittivity $\varepsilon_0$ and small anisotropic part $\hat{\nu}$:

$$\hat{\varepsilon} = \varepsilon_0 \hat{I}_3 + \hat{\nu}, \tag{2}$$

where $\hat{I}_a$ is the unit matrix of the $a$ th rank. It is convenient to characterize the weak anisotropy by small parameter

$$\mu_A = \frac{\max|\nu_{ij}|}{\varepsilon_0} \ll 1, \tag{3}$$

additional to the traditional geometrical optics parameter (1).

Solution of Maxwell equations for monochromatic wave field,

$$\text{curl}\,\text{curl}\,\mathbf{E} = k_0^2 \hat{\varepsilon} \mathbf{E}, \tag{4}$$

for the case of weakly anisotropic and smoothly inhomogeneous medium can be obtained by asymptotic expansion of the electromagnetic wave field $\mathbf{E}$ in combined small parameter

$$\mu = \max(\mu_G, \mu_A). \tag{5}$$

This results in the equations of the *quasi-isotropic approximation* (QIA) of the geometrical optics method [5–9].

In the lowest order of quasi-isotropic approximation, the monochromatic electromagnetic wave field has a form of transverse wave

$$\mathbf{E} = A\mathbf{\Gamma}\exp(ik_0\Psi - i\omega t). \tag{6}$$

In frame of QIA theory weak anisotropy influences on polarization vector $\mathbf{\Gamma}$ rather than on the eikonal $\Psi$ and amplitude $A$: the latter obey the same equations as in the isotropic medium. It means that, for lossy medium with complex dielectric permittivity $\varepsilon_0 = \varepsilon_0' + i\varepsilon_0''$, $|\varepsilon_0''/\varepsilon_0'| \ll 1$ and refractive index $n_0 = n_0' + in_0'' = \sqrt{\varepsilon_0} \simeq \sqrt{\varepsilon_0'} + i\varepsilon_0''/2\sqrt{\varepsilon_0'}$, the eikonal $\Psi$ obeys equation

$$(\nabla\Psi)^2 = \varepsilon_0', \tag{7}$$

and the amplitude $A$ satisfies transport equation

$$2\nabla A \nabla\Psi + A\Delta\Psi + k_0 \varepsilon_0'' A = 0. \tag{8}$$

In virtue of Eq. (7) the eikonal can be calculated by integrating the real part of refractive index along the ray trajectory $\mathbf{r}(\sigma)$: $\Psi = \Psi_0 + \int_0^\sigma n_0' d\sigma$, where $\sigma$ is the arc length of the ray. The rays, in turn, obey the Hamilton equations [8,9,23]

$$\dot{\mathbf{r}} = \mathbf{l}, \quad \dot{\mathbf{l}} = \nabla_\perp \ln n_0', \tag{9}$$

where the overdot stands for the derivative with respect to $\sigma$ and $\nabla_\perp = \nabla - \mathbf{l}(\mathbf{l}\nabla)$. Solution of the transport equation (8) can be presented as the solution for non-dissipative medium with additional attenuation factor $\exp\left[-k_0 \int_0^\sigma n_0'' d\sigma\right]$ [8,9].

Let us involve the unit vectors $\mathbf{e}_1$ and $\mathbf{e}_2$, which are orthogonal to each other and to the unit vector $\mathbf{l} = \dot{\mathbf{r}}$, tangent to the ray: $\mathbf{e}_1\mathbf{e}_2 = \mathbf{e}_1\mathbf{l} = \mathbf{e}_2\mathbf{l} = 0$. Polarization vector $\mathbf{\Gamma}$ of the transverse wave (6) can be expanded in the basis $(\mathbf{e}_1, \mathbf{e}_2)$ as

$$\mathbf{\Gamma} = \Gamma_1 \mathbf{e}_1 + \Gamma_2 \mathbf{e}_2, \tag{10}$$

so that $\mathbf{\Gamma l} = 0$. Longitudinal component of the polarization vector $\mathbf{\Gamma}$ appears only in the first order in small parameter $\mu$ [5–7]. In the general case the basis $(\mathbf{e}_1, \mathbf{e}_2)$ might rotate about tangent vector $\mathbf{l}$. Let $\varphi$ be the angle of rotation of the basis $(\mathbf{e}_1, \mathbf{e}_2)$ with respect to the Frenet basis $(\mathbf{n}, \mathbf{b})$, consisting of the principal normal $\mathbf{n}$ and binormal $\mathbf{b}$ to the ray, as shown at Fig. 1. Then

$$\begin{aligned}\mathbf{e}_1 &= \mathbf{n}\cos\varphi + \mathbf{b}\sin\varphi, \\ \mathbf{e}_2 &= -\mathbf{n}\sin\varphi + \mathbf{b}\cos\varphi.\end{aligned} \tag{11}$$

According to [5–7] the polarization vector (10) satisfies the following QIA equation:

$$\dot{\mathbf{\Gamma}} = \frac{i}{2}\hat{J}\mathbf{\Gamma}, \text{ with } \hat{J} = \frac{k_0}{n_0'}\begin{pmatrix} v_{11} & v_{12} + ih \\ v_{21} - ih & v_{22} \end{pmatrix}. \tag{12}$$

Here $h = 2(\kappa - \dot{\varphi})n_0'/k_0$, $\kappa$ is the ray torsion, and $v_{ab} = \mathbf{e}_a \hat{v} \mathbf{e}_b$, $a,b = 1,2$, are the "transverse" components of the tensor $\hat{v}$ in the basis $(\mathbf{e}_1, \mathbf{e}_2)$. Matrix $\hat{J}$ represents the differential Jones matrix, up to factor $i/2$ which is introduced for the convenience in what follows. The torsion $\kappa$ is defined by relation $\kappa = -\mathbf{b}\dot{\mathbf{n}}$, stemming from the Frenet-Serret formulas

$$\dot{\mathbf{l}} = K\mathbf{n}, \quad \dot{\mathbf{n}} = -K\mathbf{l} - \kappa\mathbf{b}, \quad \dot{\mathbf{b}} = \kappa\mathbf{n}. \tag{13}$$

Note that in the book [7] $\kappa$ was taken with the opposite sign, according to convention $\kappa = \mathbf{b}\dot{\mathbf{n}}$.

At $\varphi \equiv 0$ basis $(\mathbf{e}_1, \mathbf{e}_2)$, Eq. (11), coincides with the Frenet normal-binormal basis $(\mathbf{n}, \mathbf{b})$. However, the QIA equations (12) take the simplest form in frame of the "rotationless" (relative to the ray tangent $\mathbf{l}$) basis introduced first by Popov [24] to form a curvilinear orthogonal coordinate system, attached to the ray. This system is characterized by

$$\varphi = \int_0^\sigma \kappa d\sigma \tag{14}$$

and is widely used in the theory of diffraction and geophysics [25–27]. Using Frenet-Serret formulas (13) and Eq. (14) one can show that the basis vectors Eqs. (11) obey the equation

$$\dot{\mathbf{e}}_i = -(\mathbf{e}_i \nabla \ln n'_0)\mathbf{l}, \qquad (15)$$

which describes *the parallel transport* of vectors $\mathbf{e}_1$ and $\mathbf{e}_2$ along the ray [28]. According to Eq. (15), derivatives $\dot{\mathbf{e}}_i$ have not components along normal $\mathbf{n}$ and binormal $\mathbf{b}$, so that the unit vectors $\mathbf{e}_i$ do not experience rotation about the ray. Owing to Eq. (14), $h = 0$, and the Jones matrix become proportional to the "transverse" sector of the anisotropy tensor $\hat{v}$:

$$\hat{J} = \frac{k_0}{n'_0} \begin{pmatrix} v_{11} & v_{12} \\ v_{21} & v_{22} \end{pmatrix}. \qquad (16)$$

Hereafter we will use only the parallel transport coordinate frame, Eqs. (14)–(16).

For derivation of the evolution equation for the Stokes vector it is useful to write QIA equation (12) in the basis of circularly polarized waves related to the same Popov's coordinate frame. Transition to this basis is realized through substitution

$$\mathbf{\Gamma} \to \tilde{\mathbf{\Gamma}} = \hat{V}^\dagger \mathbf{\Gamma}, \quad \hat{J} \to \hat{\tilde{J}} = \hat{V}^\dagger \hat{J} \hat{V}, \text{ with } \hat{V} = \frac{1}{\sqrt{2}} \begin{pmatrix} 1 & 1 \\ i & -i \end{pmatrix}, \qquad (17)$$

in Eq. (12). Matrix $\hat{\tilde{J}}$, as well as any $2 \times 2$ matrix, can be represented as a superposition of basic unit and Pauli matrices:

$$\hat{\tilde{J}} = -G_\alpha \hat{\sigma}_\alpha, \qquad (18)$$

where

$$\hat{\sigma}_0 = \hat{I}_2, \ \hat{\sigma}_1 = \begin{pmatrix} 0 & 1 \\ 1 & 0 \end{pmatrix}, \hat{\sigma}_2 = \begin{pmatrix} 0 & -i \\ i & 0 \end{pmatrix}, \ \hat{\sigma}_3 = \begin{pmatrix} 1 & 0 \\ 0 & -1 \end{pmatrix},$$

and $G_\alpha$ is a complex 4-component vector (Greek indices take values $0,1,2,3$ and summation over repeated indices is assumed). As a result, the QIA equation (12) takes the form:

$$\dot{\tilde{\mathbf{\Gamma}}} = -\frac{i}{2} G_\alpha \hat{\sigma}_\alpha \tilde{\mathbf{\Gamma}}, \qquad (19)$$

where, in terms of the anisotropy matrix, vector $G_\alpha$ equals

$$G_0 = -\frac{k_0}{2n'_0}(v_{11} + v_{22}), \ G_1 = -\frac{k_0}{2n'_0}(v_{11} - v_{22}), \ G_2 = -\frac{k_0}{2n'_0}(v_{12} + v_{21}), \ G_3 = -i\frac{k_0}{2n'_0}(v_{12} - v_{21}). \qquad (20)$$

Real and imaginary parts of $G_\alpha$ correspond to the Hermitian and anti-Hermitian parts of the dielectric tensor (or Jones matrix) and, thus, are responsible for non-dissipative and dissipative phenomena, respectively.

## 3. Equation for the Stokes vector evolution

The 4-component Stokes vector $S_\alpha$ is defined by [23]

$$S_0 = |\Gamma_1|^2 + |\Gamma_2|^2, \ S_1 = |\Gamma_1|^2 - |\Gamma_2|^2, \ S_2 = 2\operatorname{Re}(\Gamma_1^* \Gamma_2), \ S_3 = 2\operatorname{Im}(\Gamma_1^* \Gamma_2). \qquad (21)$$

Alternatively, these expressions can be written as (use Eq. (17))

$$S_\alpha = \tilde{\mathbf{\Gamma}}^\dagger \hat{\sigma}_\alpha \tilde{\mathbf{\Gamma}}. \qquad (22)$$

By differentiating this expression and using QIA equation (19), we obtain:

$$\dot{S}_\alpha = -\frac{i}{2} \tilde{\mathbf{\Gamma}}^\dagger \left( G_\beta \hat{\sigma}_\alpha \hat{\sigma}_\beta - G_\beta^* \hat{\sigma}_\beta \hat{\sigma}_\alpha \right) \tilde{\mathbf{\Gamma}} = -\frac{i}{2} \tilde{\mathbf{\Gamma}}^\dagger \left( \operatorname{Re} G_\beta [\hat{\sigma}_\alpha, \hat{\sigma}_\beta] + i \operatorname{Im} G_\beta \{\hat{\sigma}_\alpha, \hat{\sigma}_\beta\} \right) \tilde{\mathbf{\Gamma}}, \qquad (23)$$

where square and curly brackets stand for commutators and anti-commutators, respectively. Using commutation relations for the Pauli matrices and definition (22), we arrive at equations for the evolution of Stokes vector:

$$\dot{S}_0 = \operatorname{Im} G_\beta S_\beta = \operatorname{Im} G_0 S_0 + \operatorname{Im} \mathbf{G} \mathbf{S}, \quad \dot{\mathbf{S}} = \operatorname{Re} \mathbf{G} \times \mathbf{S} + \operatorname{Im} G_0 \mathbf{S} + \operatorname{Im} \mathbf{G} S_0. \tag{24}$$

were $\mathbf{G} = (G_1, G_2, G_3)$ and $\mathbf{S} = (S_1, S_2, S_3)$ are the corresponding 3-vectors. Equations (24) can be represented in a matrix form:

$$\dot{S}_\alpha = \hat{M}_{\alpha\beta} S_\beta, \quad \hat{M} = \begin{pmatrix} \operatorname{Im} G_0 & \operatorname{Im} G_1 & \operatorname{Im} G_2 & \operatorname{Im} G_3 \\ \operatorname{Im} G_1 & \operatorname{Im} G_0 & -\operatorname{Re} G_3 & \operatorname{Re} G_2 \\ \operatorname{Im} G_2 & \operatorname{Re} G_3 & \operatorname{Im} G_0 & -\operatorname{Re} G_1 \\ \operatorname{Im} G_3 & -\operatorname{Re} G_2 & \operatorname{Re} G_1 & \operatorname{Im} G_0 \end{pmatrix}, \tag{25}$$

where $\hat{M}$ is the differential Mueller matrix for a weakly anisotropic inhomogeneous medium.

Equation (25) is the central result of this paper. One can verify that it is completely equivalent to the corresponding equation in paper by Azzam [11], but, in contrast to the latter, it is valid for arbitrary 3D smoothly inhomogeneous medium and curved rays. Transition from QIA equations to the Stokes-vector evolution equation has meaning of the transition from the Jones to Mueller calculus. As we have shown, it is realized through complex 4-component vector $G_\alpha$ which gives decomposition of the Jones matrix with respect to basic Pauli matrices, cf. [13,29–33].

Matrix $\hat{M}$ can be presented as a sum of three terms [11,20]:

$$\hat{M} = \hat{M}_a + \hat{M}_d + \hat{M}_b. \tag{26}$$

The first, *attenuation* component,

$$\hat{M}_a = \operatorname{Im} G_0 \hat{I}_4, \tag{27}$$

describes attenuation, common for all components of the Stokes vector. The second, *dichroic* term

$$\hat{M}_d = \begin{pmatrix} 0 & \operatorname{Im} G_1 & \operatorname{Im} G_2 & \operatorname{Im} G_3 \\ \operatorname{Im} G_1 & 0 & 0 & 0 \\ \operatorname{Im} G_2 & 0 & 0 & 0 \\ \operatorname{Im} G_3 & 0 & 0 & 0 \end{pmatrix} \tag{28}$$

corresponds to attenuation, responsible for dichroism, that is for selective attenuation of normal modes. At last, matrix

$$\hat{M}_b = \begin{pmatrix} 0 & 0 & 0 & 0 \\ 0 & 0 & -\operatorname{Re} G_3 & \operatorname{Re} G_2 \\ 0 & \operatorname{Re} G_3 & 0 & -\operatorname{Re} G_1 \\ 0 & -\operatorname{Re} G_2 & \operatorname{Re} G_1 & 0 \end{pmatrix} \tag{29}$$

describes *birefringence*. Note that quantity $\operatorname{Re} G_0$ does not enter in the Eq. (25), which means that the total phase of the wave is lost in the Stokes-vector evolution equation. Only the phase difference between polarization modes can be retrieved from $\operatorname{Re} G_i$.

It is worth remarking that the equation (25) for the Stokes vector evolution can be written in the form of the relativistic equation of the spin precession in an external electromagnetic field. Indeed, by assuming relativistic lowering and raising of indices (so that $S^0 = -S_0$), we have

$$\dot{S}_\alpha = F_{\alpha\beta} S^\beta + \operatorname{Im} G_0 S_\alpha, \tag{30}$$

where $S_\alpha$ plays the role of the spin 4-vector and

$$F_{\alpha\beta} = \begin{pmatrix} 0 & \operatorname{Im} G_1 & \operatorname{Im} G_2 & \operatorname{Im} G_3 \\ -\operatorname{Im} G_1 & 0 & -\operatorname{Re} G_3 & \operatorname{Re} G_2 \\ -\operatorname{Im} G_2 & \operatorname{Re} G_3 & 0 & -\operatorname{Re} G_1 \\ -\operatorname{Im} G_3 & -\operatorname{Re} G_2 & \operatorname{Re} G_1 & 0 \end{pmatrix} \equiv (\operatorname{Im} \mathbf{G}, \operatorname{Re} \mathbf{G}) \tag{31}$$

is the anti-symmetric tensor of the effective external electromagnetic field. In so doing, the imaginary and real parts of complex 3-vector **G** represent analogs of the electric and magnetic fields. Apart from the last, attenuation term, equation (30) is equivalent to the Bargmann–Michel–Telegdi equation, which follows from the Dirac equation and describes the relativistic spin precession [34–37]. This fact is a consequence of close relations between polarization optics and Lorenz group, see [29–33]. It follows from Eq. (30) that if $\operatorname{Im} G_0 = 0$, the norm of the Stokes 4-vector is conserved: $S_\alpha S^\alpha = S_1^2 + S_2^2 + S_3^2 - S_0^2 = \text{const}$; for fully polarized wave it equals zero.

For some problems it is important to know the polarization evolution on the Poincaré sphere, which is represented by the 3-component normalized Stokes vector $\mathbf{s} = \mathbf{S}/S_0$. The evolution equation for **s** can be readily obtained by differentiating its definition using Eqs. (24). As a result we arrive at

$$\dot{\mathbf{s}} = (\operatorname{Re}\mathbf{G} + \mathbf{s} \times \operatorname{Im}\mathbf{G}) \times \mathbf{s}, \quad (32)$$

which resembles the Landau–Lifshitz equation for the spin (magnetization) precession in ferromagnetic systems [38,39]. Eq. (32) ensures that the absolute value of the 3-component Stokes is conserved: $\mathbf{s}^2 = \text{const}$ (it equals to unity for a fully polarized wave). In the nondissipative medium, $\operatorname{Im} G_\alpha = 0$, equation (32) is reduced to the usual linear precession equation:

$$\dot{\mathbf{s}} = \mathbf{\Omega} \times \mathbf{s}, \quad (32')$$

where we denoted $\mathbf{\Omega} = \operatorname{Re}\mathbf{G}$. This equation also follows immediately from Eqs. (25) or (30) when $\hat{M}_a = \hat{M}_d = 0$ and $S_0 = \text{const}$. It has been derived in the general case in [22]; in paper [17] it was introduced phenomenologically in the Frenet ray coordinate frame, with an additional term proportional to the ray torsion. If $\mathbf{\Omega} = \text{const}$, the solution of equation (32') can be written as

$$\mathbf{s}(\sigma) = \hat{R}_\Omega(\Omega\sigma)\mathbf{s}(0), \quad (33)$$

where $\hat{R}_\Omega(\Omega\sigma)$ is operator of rotation on angle $\Omega\sigma$ about vector $\mathbf{\Omega}$. The fact that Stokes vector obeys the spin precession equation is quite natural, because the QIA equation (19) (apart from the diagonal term with $G_0$) has the form of the Schrödinger equation for the spin 1/2 in an external magnetic field **G**, cf. [22].

## 4. Stokes vector evolution in a magnetized plasma

In this section we will apply Eqs. (25)−(32) to analysis of the Stokes vector evolution in a weakly anisotropic plasma, considering far IR, sub-millimeter and millimeter wavelength range, typical for plasma polarimetry. In a coordinate system with $z$ axis aligned with external magnetic field $\mathbf{B}_0$, the dielectric tensor $\hat{\varepsilon}$ of a magneto-active electron plasma has the form

$$\hat{\varepsilon} = \begin{pmatrix} \varepsilon_{xx} & \varepsilon_{xy} & 0 \\ \varepsilon_{yx} & \varepsilon_{yy} & 0 \\ 0 & 0 & \varepsilon_{zz} \end{pmatrix}. \quad (34)$$

In frame of simple collisional model for electron plasma one has [1−3]

$$\varepsilon_{xx} = \varepsilon_{yy} = 1 - \frac{v(1+iw)}{(1+iw)^2 - u}, \quad \varepsilon_{zz} = 1 - \frac{v}{1+iw}, \quad \varepsilon_{xy} = -\varepsilon_{yx} = i\frac{v\sqrt{u}}{(1+iw)^2 - u}, \quad (35)$$

where $u = \left(\frac{\omega_c}{\omega}\right)^2 = \left(\frac{eB_0}{mc\omega}\right)^2$, $v = \left(\frac{\omega_p}{\omega}\right)^2 = \frac{4\pi e^2 N_e}{m\omega^2}$, and $w = \frac{v_{\text{eff}}}{\omega}$ are the standard dimensionless plasma parameters. Weak dissipation implies

$$w \ll 1, \quad (36)$$

whereas weak anisotropy, condition (3), requires
$$v \ll 1 \text{ or } u \ll 1. \tag{37}$$
Note, that only one of parameters $u$ and $v$ has to be small, while the other one can be comparable with unity. Linearization with respect to small collision parameter $w$ reduces Eqs. (35) to

$$\varepsilon_{xx} = \varepsilon_{yy} \approx 1 - \frac{v}{1-u} + i\frac{v(1+u)}{(1-u)^2}w = 1 - V + iPw, \quad \varepsilon_{zz} \approx 1 - v + ivw,$$

$$\varepsilon_{xy} = -\varepsilon_{yx} \approx i\frac{v\sqrt{u}}{1-u} + \frac{2v\sqrt{u}}{(1-u)^2}w = iV\sqrt{u} + Rw, \tag{38}$$

$$\varepsilon_{zz} - \varepsilon_{xx} \approx \frac{uv}{1-u} - i\frac{uv(3-u)}{(1-u)^2}w = uV - iQw,$$

where new dimensionless parameters are involved:
$$P = \frac{v(1+u)}{(1-u)^2}, \quad Q = \frac{uv(3-u)}{(1-u)^2}, \quad R = \frac{2v\sqrt{u}}{(1-u)^2}, \quad V = \frac{v}{1-u}. \tag{39}$$

Let $\theta$ and $\phi$ be the spherical angles indicating direction of the magnetic field $\mathbf{B}_0$ in the Popov's ray coordinate frame with orts $(\mathbf{e}_1, \mathbf{e}_2, \mathbf{l})$, Fig. 2. Then, dielectric tensor $\hat{\varepsilon}$ in this coordinate frame can be obtained by rotational transformation

$$\hat{\varepsilon} \to \hat{A}^{\dagger} \hat{\varepsilon} \hat{A}, \quad \hat{A} = \begin{pmatrix} \cos\theta\cos\phi & \cos\theta\sin\phi & -\sin\theta \\ -\sin\phi & \cos\phi & 0 \\ \sin\theta\cos\phi & \sin\theta\sin\phi & \cos\theta \end{pmatrix}. \tag{40}$$

As a result, we have the components transverse to the ray:
$$\varepsilon_{11} = \varepsilon_{xx} + (\varepsilon_{zz} - \varepsilon_{xx})\sin^2\theta\cos^2\phi, \quad \varepsilon_{12} = \varepsilon_{xy}\cos\theta + (\varepsilon_{zz} - \varepsilon_{xx})\sin^2\theta\sin\phi\cos\phi,$$
$$\varepsilon_{21} = -\varepsilon_{xy}\cos\theta + (\varepsilon_{zz} - \varepsilon_{xx})\sin^2\theta\sin\phi\cos\phi, \quad \varepsilon_{22} = \varepsilon_{xx} + (\varepsilon_{zz} - \varepsilon_{xx})\sin^2\theta\sin^2\phi. \tag{41}$$

Substitution of Eq. (38) to Eq. (41) and subtraction of the isotropic part $\varepsilon_0 = 1 - V$ from the tensor (41) yields the transverse part of the anisotropy tensor $\hat{v}$:
$$v_{11} = uV\sin^2\theta\cos^2\phi + i(P - Q\sin^2\theta\cos^2\phi)w,$$
$$v_{12} = (i\sqrt{u}V + Rw)\cos\theta + (uV - iQw)\sin^2\theta\sin\phi\cos\phi,$$
$$v_{21} = -(i\sqrt{u}V + Rw)\cos\theta + (uV - iQw)\sin^2\theta\sin\phi\cos\phi$$
$$v_{22} = uV\sin^2\theta\sin^2\phi + i(P - Q\sin^2\theta\sin^2\phi)w \tag{42}$$

From Eqs. (20) with Eq. (42) we find values $G_\alpha$:
$$G_0 = -\frac{k_0}{2n_0}\left[uV\sin^2\theta + i(2P - Q\sin^2\theta)w\right], \quad G_1 = -\frac{k_0}{2n_0}(uV - iQw)\sin^2\theta\cos 2\phi,$$
$$G_2 = -\frac{k_0}{2n_0}(uV - iQw)\sin^2\theta\sin 2\phi, \quad G_3 = \frac{k_0}{n_0}(\sqrt{u}V - iRw)\cos\theta, \tag{43}$$

where $n_0 = \sqrt{1-V}$. Finally, substituting Eq. (43) into Eqs. (25)–(29), we obtain the differential Mueller matrix for a weakly anisotropic magnetized plasma. The attenuation, dichroism, and birefringence components equal, respectively:

$$\hat{M}_a = -\frac{k_0(2P - Q\sin^2\theta)w}{2\sqrt{1-V}}\hat{I}_4, \tag{44}$$

$$\hat{M}_d = \frac{k_0 w}{2\sqrt{1-V}} \begin{pmatrix} 0 & Q\sin^2\theta\cos 2\phi & Q\sin^2\theta\sin 2\phi & -2R\cos\theta \\ Q\sin^2\theta\cos 2\phi & 0 & 0 & 0 \\ Q\sin^2\theta\sin 2\phi & 0 & 0 & 0 \\ -2R\cos\theta & 0 & 0 & 0 \end{pmatrix}, \quad (45)$$

$$\hat{M}_b = \frac{k_0 V}{2\sqrt{1-V}} \begin{pmatrix} 0 & 0 & 0 & 0 \\ 0 & 0 & -2\sqrt{u}\cos\theta & -u\sin^2\theta\sin 2\phi \\ 0 & 2\sqrt{u}\cos\theta & 0 & u\sin^2\theta\cos 2\phi \\ 0 & u\sin^2\theta\sin 2\phi & -u\sin^2\theta\cos 2\phi & 0 \end{pmatrix}. \quad (46)$$

In the case of collisionless plasma, $w=0$, when $\hat{M}_a = \hat{M}_d = 0$ and $S_0 = \text{const}$, the evolution of Stokes vector is described by the precession equation (32') with

$$\boldsymbol{\Omega} = \frac{k_0 V}{2\sqrt{1-V}} \begin{pmatrix} -u\sin^2\theta\cos 2\phi \\ -u\sin^2\theta\sin 2\phi \\ 2\sqrt{u}\cos\theta \end{pmatrix}. \quad (47)$$

Expressions (44)−(47) are similar to the Segre's results [20,21], obtained for rectilinear rays, but in contrast to the latter they are applicable to the wave propagation along curved rays with a torsion. Furthermore, some explicit distinctions are observed. In particular, the papers [20,21] do not contain the term with Q in Eq. (44) (it can be substantial when $u$ is not small) and the elements of matrix (45) is twice as large as compared with the corresponding matrix elements in [20,21] (the reason is an arithmetic inaccuracy in [20]). Note also that difference in signs in some terms is related to the sign difference in the definition of $S_3$.

## 5. Non-commutativity of the polarization evolution and conversion of normal modes

Remarkably, the Stokes-vector formalism naturally describes all non-trivial features of the polarization evolution of waves in weakly anisotropic inhomogeneous medium, such as the mutual conversion of normal modes. In paper [22] it has been associated with a non-Abelian (non-commutative) character of the polarization evolution in anisotropic inhomogeneous media, which can be readily understood by considering the non-dissipative case described by the precession equation (32'). Indeed, this equation describes $SO(3)$ evolution on the Poincaré sphere that corresponds to $U(2)$ evolution in terms of QIA equation (19) with $\text{Im}\,G_\alpha = 0$. If the medium is homogeneous, the evolution of the Stokes vector is a rotation about fixed vector $\boldsymbol{\Omega} = \text{const}$, Eq. (33), and two polarization states with stationary Stokes vectors,

$$\mathbf{s}^\pm = \pm \boldsymbol{\Omega}/\Omega, \quad (48)$$

correspond to the independent normal modes of the medium. In contrast, if the medium is inhomogeneous, vector $\boldsymbol{\Omega}$ varies along the ray, $\boldsymbol{\Omega} = \boldsymbol{\Omega}(\sigma)$, and evolution of the Stokes vector consists of $SO(3)$ rotations about different instant axes. Such rotations do not commute with one another, which evidences non-Abelian character of the evolution. There are no independent modes in this case: even if initial and final directions of $\boldsymbol{\Omega}$ coincide, the Stokes vector does not come back to its original direction. If the Stokes vector represents one of the normal modes at the input, $\mathbf{s}_\text{in} = \mathbf{s}^\pm$, the output Stokes vector, $\mathbf{s}_\text{out}$, will be different. Then, coefficient of the transformation to the other mode, opposite to $\mathbf{s}_\text{in}$, can be defined as

$$T = \frac{1 - \mathbf{s}_\text{out}\mathbf{s}_\text{in}}{2}, \quad T \in [0,1]. \quad (49)$$

Below we illustrate the non-commutative character of the polarization evolution and the mode conversion by several examples. For simplicity, we will consider discretely inhomogeneous medium consisting of few homogeneous segments. Although sharp boundaries violate condition (1), the geometrical optics approximation can be applied via matching of solutions in each homogeneous part with the help of boundary conditions at the interfaces. We suppose that the boundary conditions provide continuous evolution of the Stokes vector, i.e. it does not experience jumps at the interfaces. All further examples can be applied to smoothly inhomogeneous media as well via differential limit of the discretely-inhomogeneous medium consisting of many small segments separated by low-contrast interfaces.

First, consider propagation of wave through two homogeneous segments, characterized by vectors $\mathbf{\Omega}_I$ and $\mathbf{\Omega}_{II}$, Fig. 3a. By involving solution (33) for each homogeneous segment and supposed continuity of the Stokes vector evolution, one can determine the Stokes vector at the output of the system:

$$\mathbf{s}_{out} = \hat{R}_{\mathbf{\Omega}_{II}}(\Omega_{II}\sigma_{II}) \hat{R}_{\mathbf{\Omega}_I}(\Omega_I \sigma_I) \mathbf{s}_{in}. \tag{50}$$

Here $\sigma_I$ and $\sigma_{II}$ are the lengths of the two segments. Then, let us interchange two segments as shown in Fig. 3a. Evidently, the output Stokes vector of such system will be given by

$$\mathbf{s}'_{out} = \hat{R}_{\mathbf{\Omega}_I}(\Omega_I \sigma_I) \hat{R}_{\mathbf{\Omega}_{II}}(\Omega_{II}\sigma_{II}) \mathbf{s}_{in}. \tag{51}$$

In the general case, rotations about different axes do not commute, and $\mathbf{s}'_{out} \neq \mathbf{s}_{out}$, even if the input state was a normal mode of the first medium: $\mathbf{s}_{in} = \pm \mathbf{\Omega}_I / \Omega_I$. This indicates non-Abelian nature of the polarization evolution in weakly anisotropic inhomogeneous medium.

Second, let the medium consists of three homogeneous segments: two equivalent side segments and a distinct middle one, Fig. 3b. The Stokes vector at the output of such system equals

$$\mathbf{s}_{out} = \hat{R}_{\mathbf{\Omega}_I}(\Omega_I \sigma_I) \hat{R}_{\mathbf{\Omega}_{II}}(\Omega_{II}\sigma_{II}) \hat{R}_{\mathbf{\Omega}_I}(\Omega_I \sigma_I) \mathbf{s}_{in}. \tag{52}$$

Due to the non-commutativity of rotations, the output polarization state differs from the input one, even if the input state corresponded to normal mode of the first medium, $\mathbf{s}_{in} = \pm \mathbf{\Omega}_I / \Omega_I$. This evidences conversion of normal modes in inhomogeneous anisotropic medium. As an example of the mode conversion, let us consider magnetoactive plasma discussed in the previous Section. Supposing in-plane external magnetic field, $\phi = 0$, expression (47) yields

$$\mathbf{\Omega} = \frac{k_0 V \sqrt{u}}{2\sqrt{1-V}} \begin{pmatrix} -\sqrt{u} \sin^2 \theta \\ 0 \\ 2\cos\theta \end{pmatrix}. \tag{53}$$

Here the third component is proportional to the longitudinal magnetic field component $B_{0\|}$ and responsible for the Faraday effect, whereas the first component is proportional to the square of the transverse magnetic field component, $B_{0\perp}^2$, and corresponds to the Cotton-Mouton effect. If $u \ll 1$, the Cotton-Mouton effect is noticeable only when the magnetic field is near-orthogonal to the wave propagation direction, i.e. $\theta \approx \pi/2$ [1−4,6,7,40−42]. Let the medium depicted in Fig. 3b consists of magnetoactive plasmas with equal parameters and different directions of the external magnetic field. In the first and the third segments the Cotton-Mouton term is negligible, while in the second one $\theta = \pi/2$ and the Faraday effect vanishes, so that

$$\mathbf{\Omega}_I \approx \frac{k_0 V \sqrt{u}}{2\sqrt{1-V}} \begin{pmatrix} 0 \\ 0 \\ 2\cos\theta \end{pmatrix}, \quad \mathbf{\Omega}_{II} = \frac{k_0 V \sqrt{u}}{2\sqrt{1-V}} \begin{pmatrix} -\sqrt{u} \\ 0 \\ 0 \end{pmatrix}. \tag{54}$$

The normal modes of the Faraday medium I are circularly polarized waves, and we set $\mathbf{s}_{in} = \mathbf{s}^{\pm} = \pm(0,0,1)$. Then by applying Eq. (52) with Eq. (54), we arrive at $\mathbf{s}_{out}\mathbf{s}_{in} = \cos(\Omega_{II}\sigma_{II})$, so that the transformation coefficient (49) equals

$$T = \frac{1 - \cos\left(k_0 V u \sigma_{II} / 2\sqrt{1-V}\right)}{2}. \tag{55}$$

Finally, the polarization evolution in a weakly anisotropic medium can be non-reversible. Let the wave propagates through a homogeneous segment of magnetoactive plasma and comes back via the same path, Fig. 3c. We assume that the wave does not change its polarization state as it turns back (e.g., it can be reflected by a corner reflector). Then, the backward way is equivalent to the forward one with the inversion of the external magnetic field: $\mathbf{B}_0 \to -\mathbf{B}_0$. However, vector $\mathbf{\Omega}$, Eq. (53), is neither even nor odd function of $\mathbf{B}_0$ (the Faraday and Cotton-Mouton terms are odd and even functions, respectively). As a result, the problem is similar to the problem of the wave propagation through two *different* segments, Eq. (50), with

$$\mathbf{\Omega}_{I} = \frac{k_0 V \sqrt{u}}{2\sqrt{1-V}} \begin{pmatrix} -\sqrt{u}\sin^2\theta \\ 0 \\ 2\cos\theta \end{pmatrix}, \quad \mathbf{\Omega}_{II} = \frac{k_0 V \sqrt{u}}{2\sqrt{1-V}} \begin{pmatrix} -\sqrt{u}\sin^2\theta \\ 0 \\ -2\cos\theta \end{pmatrix}, \tag{56}$$

and $\sigma_I = \sigma_{II}$. After two rotations about different vectors (56), the output polarization state will differ from the input one: $\mathbf{s}_{out} \neq \mathbf{s}_{in}$, even if the input state was a normal mode of the medium: $\mathbf{s}_{in} = \pm\mathbf{\Omega}_I/\Omega_I$. This indicates non-reversibility of the polarization evolution and conversion of modes under double passage through the magnetoactive medium.

## 6. Conclusion

To summarize, we have derived the evolution equation for the full (four-component) Stokes vector varying as the electromagnetic wave propagates in a weakly anisotropic smoothly inhomogeneous medium. As a starting point we used equations of quasi-isotropic approximation (QIA), which follow in a consequent asymptotic way from Maxwell equations. The derived equation happens to be similar in form to the equations obtained in [10−21] for the refractionless case. However, in contrast to the prior results, it stems directly from Maxwell equations and describes the polarization evolution along curved rays in an anisotropic inhomogeneous medium of the generic type. Transition from QIA to the Stokes vector formalism (SVF) has the meaning of transition from the Jones to Mueller calculus. It is realized through complex 4-component vector which gives decomposition of the Jones matrix with respect to Pauli matrices. In these terms, the equation for the Stokes vector evolution takes an elegant form of the relativistic spin precession equation. As a result, SVF acquires all the merits of QIA. Although both approaches under consideration are equivalent (up to a total phase), SVF can be much more appropriate and beneficial in some problems. In particular, it naturally describes complex non-Abelian polarization evolution on the Poincaré sphere and mutual conversion of normal modes in a weakly anisotropic inhomogeneous medium. We have applied our general theory for analysis of the polarization evolution in a magnetized plasma and illustrated fundamental features of the polarization evolution by simple characteristic examples. Note that evolution of polarization in inhomogeneous weakly anisotropic media can also be described through an approach alternative to both QIA and SVF, namely, via the complex polarization angle [43]. Finally, in this paper we considered only polarization variations, neglecting small double-refraction effects in weakly anisotropic inhomogeneous media; the analysis of the latter can be found in [22].

*Note added.* – Recent papers [44,45] with related arguments on the Stokes-vector evolution in anisotropic media came to our attention after the submission of this work. Paper [44] discusses evolution of the three-component Stokes vector and similarity with the quantum-mechanical formalism, cf. [22]. Evolution of the four-component Stokes vector is analyzed in [45], where a similarity with the evolution of *momentum* of a relativistic particle is established.

## Acknowledgements

This work was supported in part by Association Euratom-IPPLM and Polish Ministry of Science and Higher Education (Project P-12), STCU (grant P-307), and CRDF (grant UAM2-1672-KK-06).

# References

1. K.G. Budden, *Radio Waves in the Ionosphere* (Cambridge Univ. Press, Cambridge UK 1961).
2. V.I. Ginzburg, *Propagation of Electromagnetic waves in Plasma* (Gordon and Breach, New York, 1970).
3. D.B. Melrose and R.C. McPhedran, *Electromagnetic Processes in Dispersive Media* (Cambridge Univ. Press, Cambridge UK 1991).
4. V.V. Zheleznyakov, V.V. Kocharovski, and Vl.V. Kocharovski, "Linear interaction of electromagnetic waves in inhomogeneous weakly anisotropic media", Usp. Fiz. Nauk **141**, 257-282 (1983) [Sov. Phys. Usp. **26**, 877-902 (1983)].
5. Yu.A. Kravtsov, "Quasi-isotropic geometrical optics approximation", Sov. Phys.-Doklady, **13**, 1125-1127 (1969).
6. Yu.A. Kravtsov, O.N. Naida, and A.A. Fuki, "Waves in weakly anisotropic 3D inhomogeneous media: quasi-isotropic approximation of geometrical optics", Physics-Uspekhi **39**, 129-134 (1996).
7. A.A. Fuki, Yu.A. Kravtsov, and O.N. Naida, *Geometrical Optics of Weakly Anisotropic Media* (Gordon & Breach, Lond., N.Y., 1997).
8. Yu.A. Kravtsov and Yu.I. Orlov, *Geometrical optics of inhomogeneous media* (Springer Verlag, Berlin, Heidelberg, 1990).
9. Yu.A. Kravtsov, *Geometrical Optics in Engineering Physics* (Alpha Science, London, 2005).
10. G.N. Ramachandran and S. Ramaseshan, *Crystal Optics. Encyclopedia of Physics* **25**/1 (Springer, Berlin, 1961).
11. R.M.A. Azzam, "Propagation of partially polarized light through anisotropic media with or without depolarization: a differential $4 \times 4$ matrix calculus", J. Opt. Soc. Am. **68**, 1756-1767 (1978).
12. C. Brosseau, "Evolution of the Stokes parameters in optically anisotropic media", Opt. Lett. **20**, 1221–1223 (1995).
13. C.S. Brown and A.E. Bak, "Unified formalism for polarization optics with application to polarimetry on a twisted optical fiber", Opt. Eng. **34**, 1625–1635 (1995).
14. S.E. Segre, "On the use of polarization modulation in combined interferometry and polarimetry", Plasma Phys. Contr. Fusion **40**, 153-161 (1998).
15. S.E. Segre, "A review of plasma polarimetry – theory and methods", Plasma Phys. Control. Fusion **41**, R57–R100 (1999).
16. S.E. Segre, "Evolution of the polarization state for radiation propagating in a nonuniform, birefringent, optically active and dichroic medium: the case of magnetized plasma", J. Opt. Soc. Am. A **17**, 95–100 (2000).
17. S.E. Segre, "Effect of ray refraction in evolution of the polarization state of radiation propagating in a nonuniform, birefringent, optically active and dichroic medium", J. Opt. Soc. Am. A **17**, 1682–1683 (2000).
18. S.E. Segre, "New formalism for the analysis of polarization evolution for radiation in a weakly nonuniform, fully anisotropic medium: a magnetized plasma", J. Opt. Soc. Am. A **18**, 2601-2606 (2001).
19. S.E. Segre, "Comparison between two alternative approaches for the analysis of polarization evolution of EM waves in a nonuniform, fully anisotropic medium: a magnetized plasma", Preprint RT/ERG/FUS/2001/13, ENEA, Frascati, Italy (2001).
20. S.E. Segre, "Polarization evolution of radiation propagating in weakly non-uniform magnetized plasma with dissipation", J. Phys. D: Appl. Phys. **36**, 2806–2810 (2003).
21. S.E. Segre and V. Zanza, "Derivation of the pure Faraday and Cotton–Mouton effects when polarimetric effects in a tokamak are large", Plasma Phys. Control. Fusion **48**, 339–351 (2006).

22. K.Yu. Bliokh, D.Yu. Frolov, and Yu.A. Kravtsov, "Non-Abelian evolution of electromagnetic waves in a weakly anisotropic inhomogeneous medium", Phys. Rev. A **75**, 053821 (2007).
23. M. Born and E. Wolf, *Principles of Optics* (Pergamon, Oxford, 1980).
24. M.M. Popov, "Eigen-oscillations of multi-mirrors resonators", Vestnik Leningradskogo Universiteta **22**, 44-54 (1969) [in Russian]. Derivation of the Popov's orthogonal coordinate system is reproduced also in the Ch. 9 of the book [24].
25. V.M. Babich and V.S. Buldyrev, *Short-Wavelength Diffraction Theory: Asymptotic Methods* (Springer Verlag, Berlin, 1990) [Original Russian edition: V.M. Babich and V.S. Buldyrev. *Asymptotic Methods in Short-Wavelength Diffraction Problems: The Model Problem Method* (Nauka, Moscow, 1972)].
26. V. Červený, *Seismic Ray Theory* (Cambridge University Press, 2001).
27. P. Berczynski, K.Yu. Bliokh, Yu.A. Kravtsov, and A. Stateczny, "Diffraction of Gaussian beam in 3D smoothly inhomogeneous media: eikonal-based complex geometrical optics approach", J. Opt. Soc. Am. A **23**, 1442-1451 (2006).
28. For review see S.I. Vinitsky, V.L. Debrov, V.M. Dubovik, B.L. Markovski, and Yu.P. Stepanovskii, "Topological phases in quantum mechanics and polarization optics", Usp. Fiz. Nauk **160**(6), 1–49 (1990) [Sov. Phys. Usp. **33**, 403–450 (1990)].
29. R. Barakat, "Bilinear constraints between elements of the 4x4 Mueller-Jones transfer matrix of polarization theory", Opt. Comm. **38**, 159–161 (1981).
30. R. Simon, "The connection between Mueller and Jones matrices of polarization optics", Opt. Comm. **42**, 293–297 (1982).
31. S.R. Cloude, "Group theory and polarization algebra", Optik **75**, 26–36 (1986).
32. K. Kim, L. Mandel, and E. Wolf, "Relationship between Jones and Mueller matrices for random media", J. Opt. Soc. Am. A **4**, 433–437 (1987).
33. D. Han, Y.S. Kim, and M.E. Noz, "Polarization optics and bilinear representation of the Lorentz group", Phys. Lett. A **219**, 26–32 (1996).
34. V.B. Berestetskii, E.M. Lifshits, and L.P. Pitaevskii, *Quantum Electrodynamics* (Pergamon, Oxford, 1982).
35. V. Bargmann, L. Michel, and V.L. Telegdi, "Precession of the polarization of particles moving in a homogeneous electromagnetic field" Phys. Rev. Lett. **2**, 435–436 (1959).
36. J. Bolte and S. Keppeler, "A semiclassical approach to the Dirac equation", Ann. Phys. (NY) **274**, 125–162 (1999).
37. H. Spohn, "Semiclassical limit of the Dirac equation and spin precession" Ann. Phys. (NY) **282**, 420–431 (2000).
38. A.I. Akhiezer, V.G. Baryakhtar, and S.V. Peletminskii, *Spin Waves* (North-Holland, Amsterdam, 1968).
39. C.P. Slichter, *Principles of Magnetic Resonance* (Springer-Verlag, New York, 1989).
40. Yu.A. Kravtsov and O.N. Naida, "Linear transformation of electromagnetic waves in three-dimensional inhomogeneous magneto-active plasma", Sov. Phys. - JETP **44**, 122-126 (1976).
41. Yu.A. Kravtsov and O.N. Naida, "Theory for Cotton-Mouton diagnostics of magnetized plasma", J. Tech. Phys. (Poland) **41**, 155-160 (2000).
42. Yu.A. Kravtsov and O.N. Naida, "Manifestation of the Cotton-Mouton effect in the ionosphere plasma", Advanced Space Research **27**, 1233-1237 (2001).
43. Z.H. Czyz, B. Bieg, and Yu.A. Kravtsov, "Complex polarization angle: relation to traditional polarization parameters and application to microwave plasma polarimetry" Phys. Lett. A (2007, in press).
44. H. Kuratsuji and S. Kakigi, "Maxwell-Schrödinger equation for polarized light and evolution of the Stokes parameters", Phys. Rev. Lett. **80**, 1888–1891 (1998).
45. R. Botet, H. Kuratsuji, and R. Seto, "Novel aspects of evolution of the Stokes parameters for an electromagnetic wave in anisotropic media", Prog. Theor. Phys. **116**, 285–294 (2006).

**Figures captions**

**Fig. 1.** Coordinate frames locally attached to the ray (dotted curve): Popov's (parallel transport) basis $(\mathbf{e}_1, \mathbf{e}_2, \mathbf{l})$ vs. Frenet (natural trihedral) basis $(\mathbf{n}, \mathbf{b}, \mathbf{l})$.

**Fig. 2.** External magnetic field $\mathbf{B}_0$ in the Popov's coordinate frame. Longitudinal and transverse (with respect to the ray) components are given by $\mathbf{B}_{0\|} = \mathbf{l}(\mathbf{B}_0 \mathbf{l})$ and $\mathbf{B}_{0\perp} = \mathbf{B}_0 - \mathbf{l}(\mathbf{B}_0 \mathbf{l})$.

**Fig. 3.** Schematic illustration of the peculiarities of the polarization evolution in weakly anisotropic inhomogeneous media: (a) – non-commutativity, (b) conversion of modes, (c) non-reversibility.

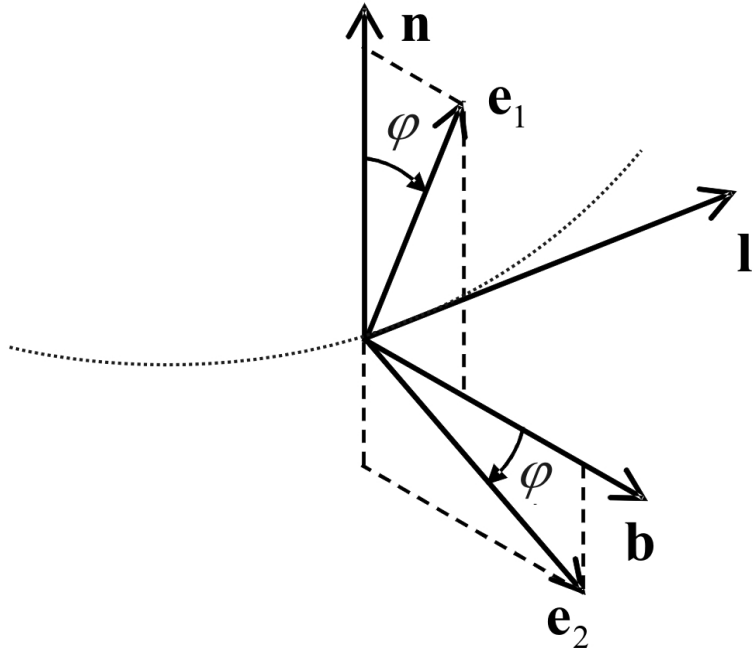

**Fig. 1**

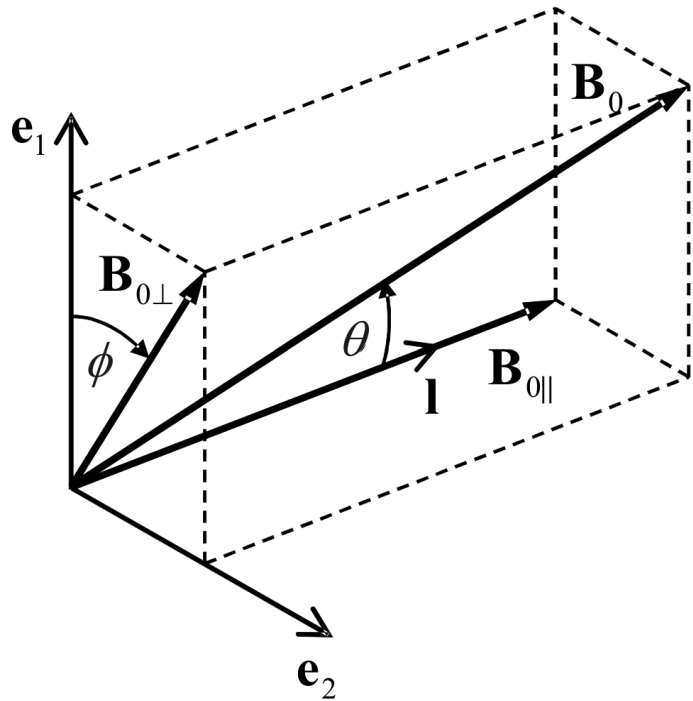

**Fig. 2**

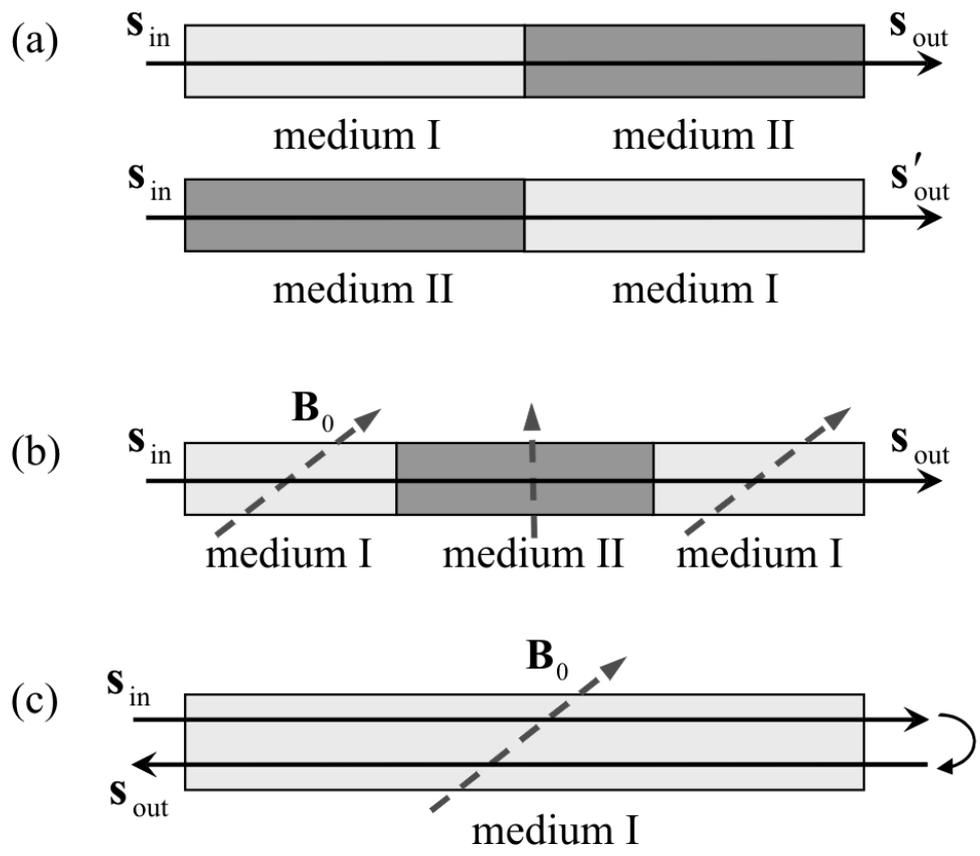

**Fig. 3**